\documentclass[twocolumn,showpacs,preprintnumbers,amsmath,amssymb]{revtex4}

\usepackage{graphicx}% Include figurthebibliographye files
\usepackage{dcolumn}% Align table columns on decimal point
\usepackage{bm}% bold math

\begin{document}
\preprint{COAA/ON - CG09 - 07/15/09}
%\preprint{DFTE/UFRN -  - 07/15/09}
\def\be{\begin{equation}}
\def\ee{\end{equation}}
\def\lb{\label}

\title{Nonextensive Quantum H-Theorem}

%%%%%%%%%%%%%%%%%%%%%%%%%%%%%%%%%
\author{R. Silva$^{1,2}$} \email{raimundosilva@dfte.ufrn.br}

\author{D. H. A. L. Anselmo$^{1}$} \email{doryh@dfte.ufrn.br}

\author{J. S. Alcaniz$^{3}$} \email{alcaniz@on.br}

\affiliation{$^{1}$Universidade Federal do Rio Grande do Norte,
Departamento de F\'{\i}sica, Natal-RN, 59072-970, Brasil}

\affiliation{$^{2}$Universidade do Estado do Rio Grande do Norte,
Departamento de F\'{\i}sica, Mossor\'o-RN, 59610-210, Brasil}

\affiliation{$^{3}$Departamento de Astronomia, Observat\'orio
Nacional, 20921-400, Rio de Janeiro-RJ, Brasil}

\date{\today}

\begin{abstract}

A proof of the quantum $H$-theorem taking into account nonextensive
effects on the quantum entropy $S^Q_q$ is shown. The positiveness of
the time variation of $S^Q_q$ combined with a duality transformation
implies that the nonextensive parameter $q$ lies in the interval
[0,2]. It is also shown that the equilibrium states are described by
quantum $q$-power law extensions of the Fermi-Dirac and Bose-Einstein
distributions. Such results reduce to the standard ones in the
extensive limit, thereby showing that the nonextensive entropic
framework can be harmonized with the quantum distributions contained
in the quantum statistics theory.

\end{abstract}

\pacs{05.30.-d; 03.65.Ta; 05.70.Ln}

\keywords{Non-extensive statistical mechanics, fermions, bosons, H-theorem.}

\maketitle

\section{Introduction}

The Boltzmann's famous $H$-theorem, which guarantees positive-definite
entropy production outside equilibrium, also describes the increase in
the entropy of an ideal gas in an irreversible process, by considering
the Boltzmann equation. Roughly speaking, this theorem implies that in
the thermodynamical equilibrium the distribution function of an ideal
gas evolves irreversibly toward Maxwellian equilibrium distribution
\cite{Tolman}. In the special relativistic domain, the very fisrt
derivation of this theorem was done by Marrot \cite{2} and, in the
local form, by Ehlers \cite{3}, Tauber and Weinberg \cite{4} and
Chernikov \cite{5}. As well known, the $H$-theorem furnishes the
Juttner distribution function for a relativistic gas in equilibrium,
which contains the number density, the temperature, and the local
four-momentum as free parameters \cite{6}. In the quantum domain, the
first derivation was done by Pauli \cite{pauli28}, which showed that
the change of entropy with time as a result of collisional
equilibrium states are described by Bose-Einstein and Fermi-Dirac
distributions.

Recently, a considerable effort has been done toward the development
of a generalization of thermodynamics and statistical mechanics aiming
at better understanding a number of physical systems that possess
exotic properties, such as broken ergodicity, strong correlation
between elements, multi fractality of phase-space and long-range
interactions. In this regard, the nonextensive statistical mechanics
(NESM) framework proposed by Tsallis~\cite{TG} is based on the
nonadditive $q$-entropy
\begin{equation}\label{1}
S_q=k{1-\sum_{i=1}^W p_i^q \over q-1},
\end{equation}
where $k$ is a positive constant, $W$ is the number of microscopic
states, and $p_i$ is a normalized probability distribution. In this
approach, additivity for two probabilistically independent subsystems
A and B is generalized by the following pseudo-additivity:
\be
{S_q (\rm{A , B})} = {S_q (\rm{A})} + {S_q (\rm{B})} + (1 - q)
\frac{S_q (\rm{A}) S_q (\rm{B})}{k}\;.
\ee
For subsystems that have special probability correlations, extensivity
may be no longer valid, so that a more realistic description may be
provided by the $S_q$ form with a particular value of the index $q\neq
1$, called the $q$-entropic parameter. In the limit $q \rightarrow 1$,
not only the Boltzmann-Gibbs (BG) entropy $S_1=k\sum_{i=1}^W p_i\ln
p_i$ is fully recovered, but so is the additivity property for the
subsystems A and B above, i.e., $S_{\rm{BG}}(\rm{A,B})=S_{\rm{BG}}
(\rm{A})+S_{\rm{BG}}(\rm{B})$. \par Several consequences of this generalized framework have been
investigated in the literature~\cite{q-application} and we refer the
reader to Ref.~\cite{update} for a regularly updated bibliography. In
particular, it is worth mentioning that the proofs of both the
non-relativistic and relativistic nonextensive $H$-theorem have been discussed in Refs.~\cite{prl2001,pre05,ths} 
%(For others approachs, see Ref. \cite{ths}).

The aim of this paper is twofold. First, to derive a proof of the
quantum $H$-theorem by including nonextensive effects on the quantum
entropy $S^Q$ in the Tsallis formalism. %($dS_q^Q/dt \geq 0$).
Second, to obtain from this proof a natural generalization of the
quantum Bose-Einstein and Fermi-Dirac distributions. It is shown that
the equilibrium states are simply described by a $q$-power law
extension of the usual Fermi-Dirac and Bose-Einstein distributions. From the positiveness of the rate $dS_q^Q/dt$ we also discuss possible
constraints on the dimensionless index $q$.

\section{Quantum H-Theorem}

Let us start by presenting the mains results of the standard
$H$-theorem in quantum statistical mechanics. The first one is a
specific functional form for the entropy~\footnote{In this context, we
assume a gas appropriately specified by regarding the states of energy
for a single particle in the container as divided up into groups of
$g_\kappa$ neighboring states, and by stating the number of particles
$n_\kappa$ assigned to each such group $\kappa$.}, which is expressed
by logarithmic measure \cite{Tolman}.

\be \label{SQ}
S^Q = -\sum_\kappa [n_\kappa \ln n_\kappa \mp (g_\kappa \pm n_\kappa)
\ln (g_\kappa \pm n_\kappa)\pm g_\kappa \ln g_\kappa].
\ee

The second one is the well-known expression for quantum distributions,
which is the rule of counting of quantum states in the case of
Bose-Einstein and Fermi-Dirac gases
\be
n_\kappa={g_\kappa\over e^{\alpha+\beta\epsilon_\kappa}\mp 1}.
\ee
These two statistical expressions are the pillars of the quantum
$H$-theorem. As is well known, the evolution of $S$ with time as a
result of molecular collisions leads to the quantum distributions
$n_\kappa$.

\subsection{Proof of $H_q$-theorem}

Let us now consider a spatially homogeneous gas of $N$ particles
(bosons or fermions) enclosed in a volume $V$. In this case, the time
derivative of the particle number $n_\kappa$ is given by considering
collisions of pairs of particles, where a pair of particles goes from
a group $\kappa,\lambda$ to another group $\mu, \nu$. Here, the
expected number of collisions per unit of time is given
by~\footnote{In other words, the collisions in the sample of gas in a condition
specified by taking $n_\kappa, n_\lambda, n_\mu, n_\nu, ...$ as the
numbers of particles in different possible groups of
$g_\kappa,g_\lambda,g_\mu,g_\nu, ...$, elementary states, are described quantitatively by
$Z_{\kappa \lambda,\mu \nu}$(For details see Ref.~\cite{Tolman}).}
\be
Z_{\kappa \lambda,\mu \nu} =  A_{\kappa \lambda,\mu \nu}n_\kappa
n_\lambda(g_\mu \pm n_{\mu})(g_\nu \pm n_{\nu})\;,
\ee
where, as before, the upper sign refers to bosons, and the lower one
to fermions. The coefficient $A_{\kappa \lambda,\mu \nu}$ must satisfy
the relation
\be
\label{coefficients}
A_{\kappa \lambda,\mu \nu} = A_{\mu \nu, \kappa \lambda}\;,
\ee
which in turn determines the frequency of collisions that are inverse
to those considered, i.e., collisions in which particles are thrown
from $\mu,\nu$ to $\kappa,\lambda$ instead of from $\kappa,\lambda$ to
$\mu,\nu$.
This coefficient must have a value close to zero for collisions which
do not satisfy the energy partition:
\be
\label{energy}
\epsilon_{\mu}  + \epsilon_{\nu}  = \epsilon_{\kappa}  + \epsilon_{\lambda}\;.
\ee
With the assumptions above, the time derivative of $n_\kappa$ reads
%\begin{widetext}
\begin{eqnarray}
\label{dn/dt}
\frac{dn_\kappa}{dt} & = & - \sum_{\lambda, (\mu \nu)} A_{\kappa
\lambda,\mu \nu} n_\kappa n_{\lambda}(g_\mu \pm n_{\mu})(g_\nu \pm
n_{\nu})\\ \nonumber & & + \sum_{\lambda, (\mu \nu)} A_{\mu \nu,
\kappa \lambda} n_{\mu} n_{\nu}(g_\kappa \pm n_{\kappa})(g_\lambda \pm
n_{\lambda})\;,
\end{eqnarray}
%\end{widetext}
where the sum above spans over all groups $\lambda$ and also over all
pairs of groups ($\mu \nu$). Also, we make a double inclusion of those
terms in the summation for which $\lambda = \kappa$.

In order to study the influence of the NESM on the quantum
$H$-theorem, we first introduce the generalized entropic measure
defined in Ref.~\cite{Sq}, i.e.,
\begin{equation}
\label{Sq1}
S_q^Q = - \sum_\kappa n_\kappa^q \ln_q n_\kappa \mp (g_\kappa \pm
n_\kappa)^q \ln_q (g_\kappa \pm n_\kappa) \pm g_\kappa^q\ln_q
g_\kappa,
\end{equation}
where we use the functionals $H_q=-{S^Q_q / k}$.
The generalized $q$-logarithm is defined by~\cite{TG}
\be
\ln_q(x) := \frac{x^{1-q} - 1}{1-q}\;
\ee
whose inverse function is given by the $q$-exponential function
\be \label{eqx}
\exp_q(x) := [1 - (1-q)x]^{1/(1-q)}\;.
\ee
Note that, when $q \rightarrow 1$, Eq. (\ref{Sq1}) reduces to the
standard case (\ref{SQ}).

%When $q \rightarrow 1$, Eq. (\ref{Sq1}) reduces to quantum entropy.

By taking the time derivative of $S_q^Q$, we obtain
\be \label{sq0}
\frac{dS_q^Q}{dt}= -q \sum_\kappa [{\ln_{q^*} n_\kappa - \ln_{q^*}
(g_\kappa \pm n_\kappa)}]\frac{dn_\kappa}{dt}\;,
\ee
where we have used the transformation $f^{q-1} \ln_q f = \ln_{q*} f$
with $q*= 2-q$. Now, we make use of the so-called $q$-algebra,
introduced in Ref. \cite{borges}, and define the $q$-difference and the $q$-product, respectively, as
\begin{subequations}
\be \label{qdif}
x \ominus_{q*} y := \frac{x-y}{1+(1-q^*)y} \ \ \ \ \forall \ \ y \neq \frac{1}{1-q^*},
\ee
\be \label{qprod}
x \otimes_{q*} y := \left[x^{1-q_*} + y^{1-q_*} - 1 \right]^{\frac{1}{1-q_*}} \ \ \ \ x, y >0
\ee
\end{subequations}
and the $\ln_q$ of a product and a of quotient
\begin{subequations}
\be \label{lnqprod}
\ln_{q^*}({x} \otimes_{q*} {y} ) := \ln_{q^*} (x) + \ln_{q^*} (y)\;,
\ee
\be \label{lnqquot}
\ln_{q^*} (x) \ominus_{q*} \ln_{q^*} (y) := \ln_{q^*}(\frac{x}{y} )\;.
\ee
\end{subequations}

From definitions (\ref{qdif}) and (\ref{lnqquot}), we can rewrite the term in square brackets
in Eq. (\ref{sq0}) as
\be
\ln_{q^*}(\frac{n_\kappa}{g_\kappa \pm n_\kappa}) = \frac{\ln_{q^*}
n_\kappa - \ln_{q^*} (g_\kappa \pm n_\kappa)}{\widetilde{n}_\kappa}\;,
%1+(1-q)\ln_{q^*}(g_\kappa \pm n_\kappa)}\;,
\ee
so that Eq. (\ref{sq0}) reads
\be \label{sq}
\frac{dS_q^Q}{dt}= q \sum_\kappa [\ln_{q^*}(\frac{n_\kappa}{g_\kappa
\pm n_\kappa} )] \cdot  \widetilde{n}_\kappa \frac{dn_\kappa}{dt}\;,
\ee
where
\be
\widetilde{n}_\kappa = 1+(1-q*)\ln_{q^*}(g_\kappa \pm n_\kappa) =
2-(g_\kappa \pm n_\kappa)^{q*-1}\;.
\ee

Substituting (\ref{dn/dt}) into (\ref{sq}), we arrive at
\begin{widetext}
\begin{eqnarray}
\frac{dS_q^Q}{dt}= q\sum_\kappa \sum_{\lambda,(\mu \nu)} A_{\kappa
\lambda,\mu \nu} n_\kappa n_{\lambda}\widetilde{n}_\kappa (g_\mu \pm
n_{\mu})(g_\nu \pm n_{\nu})\cdot \ln_{q^*}(\frac{n_\kappa}{g_\kappa
\pm n_\kappa}) - &&\\ - q\sum_\kappa \sum_{\lambda,(\mu \nu)} A_{\mu
\nu, \kappa \lambda} n_{\mu} n_{\nu} \widetilde{n}_\kappa(g_\kappa \pm
n_\kappa)(g_\lambda \pm n_{\lambda}) \cdot
\ln_{q^*}(\frac{n_\kappa}{g_\kappa \pm n_\kappa})
\end{eqnarray}
\end{widetext}
where the summations include all groups $\kappa$ and $\lambda$ and all
pairs of groups ($\mu \nu$).

In order to rewrite $dS_q^Q/dt$ in a more symmetrical form some
elementary operations must be done in the above expression. Following
standard lines [1], we first notice that changing to a summation over
all pairs of groups ($\kappa,\lambda$) does not affect the value of
the sum. This happens because the coefficients satisfies the equality
for inverse collisions [see Eq. (\ref{coefficients})]. By implementing
these operations and symmetrizing the resulting expression, $dS_q^Q/dt$
can be rewritten as
\begin{widetext}
\begin{eqnarray}
\frac{dS_q^Q}{dt}& = & {q\over 2}\sum_{(\kappa \lambda),(\mu \nu)}
A_{\kappa \lambda,\mu \nu} \widetilde{n}_\kappa \widetilde{n}_\lambda
(g_\mu \pm n_{\mu})(g_\nu \pm n_{\nu})(g_\kappa \pm
n_{\kappa})(g_\lambda \pm n_{\lambda}) \times \\ && \times
\left[\frac{n_\kappa}{g_\kappa \pm
n_\kappa}\frac{n_{\lambda}}{g_\lambda \pm
n_\lambda}-\frac{n_{\mu}}{g_\mu \pm n_{\mu}}\frac{n_{\nu}}{g_\nu \pm
n_{\nu}}\right] \nonumber \times \\ && \times \left[
\ln_{q^*}\frac{n_\kappa}{g_\kappa \pm
n_\kappa}+\ln_{q^*}\frac{n_{\lambda}}{g_\lambda \pm n_\lambda} -
\ln_{q^*}\frac{n_{\mu}}{g_\mu \pm
n_{\mu}}-\ln_{q^*}\frac{n_{\nu}}{g_\nu \pm n_{\nu}}
\right] \nonumber
\label{sq2}
\end{eqnarray}
\end{widetext}
Note that the summation in the above equation is never negative,
because the terms $\widetilde{n}_{\kappa}$, $\widetilde{n}_{\lambda}$
and $g_j\pm n_j$ with $j=\mu,\nu,\kappa,\lambda$ are always positive
and $g_j \geq n_j$ on account for the Pauli exclusion principle. Note
also that by defining
\begin{subequations}
\be
x := \frac{n_\kappa}{g_\kappa \pm n_\kappa}\;,
\ee
\be
y := \frac{n_{\lambda}}{g_\lambda \pm n_\lambda}\;,
\ee
\be
z := \frac{n_{\mu}}{g_\mu \pm n_{\mu}}\;,
\ee
\mbox{and}
\be
w := \frac{n_{\nu}}{g_\nu \pm n_{\nu}}\;,
\ee
\end{subequations}
we can show that the function
\be
\varphi(x,y,z,w) = {\rm{X}}\cdot(\ln_{q^*}x+\ln_{q^*}y-\ln_{q^*}z-\ln_{q^*}w)\;,
\ee
where ${\rm{X}} = (xy-zw)$, is also a positive quantity.

Finally, we note that, for positive values of $q$, and by considering
the duality transformation $q^*=2-q$, i.e., $q<2$ (as pointed out in
Ref. \cite{karlin}), we obtain the quantum $H_q$-theorem~\footnote{It
is worth emphasizing that this same interval is also obtained in both
non-relativistic and relativistic regimes. See, e.g.,
\cite{prl2001}.}
\be \label{ht}
\frac{dS_q^Q}{dt} \geq 0
\ee
Note that, when $q < 0$ or $q>2$, the quantum $q$-entropy is a
decreasing function of time. Consequently, it seems that within the
present context
the parameter $q$ should be restricted to interval [0,2]. Notice also
that the entropy does not change with time if $q=0$. It should be
emphasized that, in {\it quantum regime}, the equivalent constraint on
the nonextensive parameter was also calculated based on the second law
of thermodynamics, i.e., through Clausius' inequality~\cite{abePRL03}.

In order to finalize the proof of the quantum $H$-theorem, let us now
calculate the nonextensive Fermi-Dirac and Bose-Einstein
distributions. As happens in the extensive case, $dS_q^Q/dt=0$ is a
necessary and sufficient condition for local and global equilibrium. From Eq. (\ref{sq2}), we note that two conditions must be satisfied, i.e.,
%\begin{widetext}
\begin{subequations} \label{24}
\be
\frac{n_\kappa}{g_\kappa \pm n_\kappa}\frac{n_{\lambda}}{g_\lambda \pm
n_\lambda}=\frac{n_{\mu}}{g_\mu \pm n_{\mu}}\frac{n_{\nu}}{g_\nu \pm
n_{\nu}}\;,
\ee
and
\begin{eqnarray}\label{24b}
\ln_{q^*}\frac{n_\kappa}{g_\kappa \pm
n_\kappa} & + & \ln_{q^*}\frac{n_{\lambda}}{g_\lambda \pm n_\lambda} = \\ & & \nonumber
\ln_{q^*}\frac{n_{\mu}}{g_\mu \pm
n_{\mu}}+\ln_{q^*}\frac{n_{\nu}}{g_\nu \pm n_{\nu}}\;.
\end{eqnarray}
\end{subequations}
Note also that a complete symmetry between these two latter expressions can be achieved by using definitions (\ref{qprod}) and (\ref{lnqprod}), so that Eq. (\ref{24b}) can be rewritten as
\be \label{25}
\frac{n_\kappa}{g_\kappa \pm
n_\kappa}\otimes_{q*}\frac{n_{\lambda}}{g_\lambda \pm
n_\lambda}=\frac{n_{\mu}}{g_\mu \pm
n_{\mu}}\otimes_{q*}\frac{n_{\nu}}{g_\nu \pm n_{\nu}}\;.
\ee
%\end{widetext}
Therefore, the usual product and $q$-product must be satisfied simultaneously in order to null the rate of change of $q$-quantum entropy. In particular, for a null value of this rate of change, Eq. (\ref{24b})
%\begin{eqnarray}
%\ln_{q^*}\frac{n_\kappa}{g_\kappa \pm n_\kappa} & + &
%\ln_{q^*}\frac{n_{\lambda}}{g_\lambda \pm n_\lambda} = \\ & &
%\ln_{q^*}\frac{n_{\mu}}{g_\mu \pm
%n_{\mu}}+\ln_{q^*}\frac{n_{\nu}}{g_\nu \pm n_{\nu}} \nonumber
%\end{eqnarray}
satisfies the energy relation (\ref{energy}) for collisions with appreciable value of $A_{\kappa\lambda,\mu\nu}$. Here, the above sum
of $q$-logarithms remains constant during a collision, i.e., it is a
summational invariant. In the quantum regime, the solution of these
equations is an expression of the form
\be\label{good}
\ln_{q*} {n_\kappa\over g_\kappa\pm n_\kappa} +\alpha+\beta\epsilon_\kappa=0,
\ee
where $\alpha$ and $\beta$ are constants independent of $\kappa$.
After some algebra, we may rewrite Eq.~(\ref{good}) as the quantum
nonextensive distribution
\be \label{nk}
n_\kappa = {g_\kappa \over {\exp_{q*}(\alpha+\beta\epsilon_\kappa)} \pm 1}\;,
\ee
where $\exp_{q*}(x)$ is the $q$-exponential function defined in Eq.
(\ref{eqx}). The above expression, which coincides with the
$q$-quantum distribution for fermions derived in Ref.~\cite{TPM}, seems to be the
most general expression which leads to a vanishing rate of change, and
clearly reduces to Fermi-Dirac and Bose-Einstein distribution in the
extensive limit $q\rightarrow 1$.

\section{Final remarks}

In this paper, we have investigated a $q$-generalization of the quantum $H$-theorem based on the Tsallis nonextensive thermostatistics. We have showed that the $q$-thermostatistics can be extended in order to achieve the quantum distributions concepts of the quantum statistical mechanics. In addition, their generalization to the relativistic framework can be readily accomplished.

 It should be emphasized that the combination of the quantum $H_q$-theorem and duality transformation \cite{karlin} has constrained the nonextensive parameter to interval of validity $q \in [0,2]$, which is fully consistent with the results of Refs.~\cite{abePRL03,kaniadakis2005} and also with the bounds obtained from several independent studies involving the Tsallis nonextensive framework (see, e.g. \cite{20}). In particular, for $g_\kappa=1$ and the Fermi-Dirac case, the quantum nonextensive distributions [Eq. (\ref{nk})], reproduces the result originally obtained in Ref.~\cite{TPM}. 

Finally, it is worth emphasizing that this work seems to complement a series of investigations on the compatibility between nonextensivity and the Boltzmann $H$-theorem and shows, together with Refs.~\cite{prl2001,pre05,ths}, that a nonextensive $H_q$-theorem can be derived in non-relativistic, relativistic and quantum regimes. Also, our formalism is very general, since our assumptions can be applied to any ensemble whose quantity $S_q^Q$ can be defined and calculated within the framework of Tsallis statistics, and whose value of equilibrium (steady state) is obtained by allowing this system to evolve in time.

\begin{acknowledgments}

RS would like to thank the hospitality of the Departamento de
Astronomia of Observat\'orio Nacional/MCT where part of this work was
developed. RS and JSA thank CNPq - Brazil for the grants under which
this work was carried out. DHALA acknowledges financial support from
Funda\c{c}\~ao de Amparo \`a Pesquisa do Estado do Rio Grande do Norte
- FAPERN.

\end{acknowledgments}


\begin{thebibliography}{30}
\bibitem{Sq} This is a $q$-quantum entropy which generalize the standart one. By considering the fermions and $g_\kappa=1$ this expression provides the equation of Ref. \cite{TPM}.
\end{thebibliography}
\end{document}